\documentclass[10pt, twocolumn, nofootinbib,superscriptaddress]{revtex4-2}
\usepackage{amsmath,amssymb,amsfonts}
\usepackage{algorithmic}
\usepackage{graphicx}
\usepackage{textcomp}
\usepackage{xcolor}
\usepackage{ragged2e}
\usepackage{booktabs, makecell, tabularx}
\usepackage{lipsum}

\usepackage{gensymb}
\usepackage{lineno}
\makeatletter
    \renewcommand\@make@capt@title[2]{%
     \@ifx@empty\float@link{\@firstofone}{\expandafter\href\expandafter{\float@link}}%
      {\textbf{#1}}\@caption@fignum@sep#2\quad}%
\makeatother
\makeatletter 
\newcommand{\SiN}{Si\textsubscript{3}N\textsubscript{4}}
\renewcommand{\fnum@figure}{\textbf{Fig.~\thefigure}} 
\makeatother

\usepackage{xcolor}

\def\BibTeX{{\rm B\kern-.05em{\sc i\kern-.025em b}\kern-.08em
    T\kern-.1667em\lower.7ex\hbox{E}\kern-.125emX}}

\begin{document}

\author{Riley te Morsche}
\thanks{These authors contributed equally to this work}
\author{Ahmet Tarik Işık}
\thanks{These authors contributed equally to this work}
\author{Job Noordkamp}
\author{Jitse Boonstra}
\author{Akshay Keloth}
\author{Roel Botter}
\author{Akhileshwar Mishra}
\author{Okky Daulay}
\author{Peter van der Slot}
\affiliation{Nonlinear Nanophotonics Group, MESA+ Institute of Nanotechnology,\\
University of Twente, Enschede, Netherlands}
\author{Arjan Meijerink}
\author{Edwin Klein}
\affiliation{LioniX International BV, Hengelosestraat 527, Enschede, Netherlands}
\author{Yvan Klaver}
\affiliation{Nonlinear Nanophotonics Group, MESA+ Institute of Nanotechnology,\\
University of Twente, Enschede, Netherlands}
\affiliation{Current address: Department of Microtechnology and Nanoscience (MC2), Chalmers University of Technology, Göteborg, Sweden}
\author{David Marpaung}
\email{david.marpaung@utwente.nl}
\affiliation{Nonlinear Nanophotonics Group, MESA+ Institute of Nanotechnology,\\
University of Twente, Enschede, Netherlands}

\date{\today}
\title{High-power Brillouin photonics in a low-loss, foundry Si\textsubscript{3}N\textsubscript{4} platform}

\begin{abstract}
Stimulated Brillouin scattering (SBS) is emerging as a key nonlinearity used to implement an array of unique photonic functionalities. Silicon nitride (Si\textsubscript{3}N\textsubscript{4}) is a cornerstone platform to realise demanding technologies like microwave photonics and narrow-linewidth lasers. However, generating significant SBS gain in \SiN~has remained elusive, restricting its use to Brillouin lasers in ultra-high-Q resonators or requiring heterogeneous integration. This prevents integration with more mature integrated photonics components, blocking the pathway to SBS-based devices with performance and signal-to-noise required for real-world application. Here, we introduce seamless integration of SBS-active waveguides in zero-change, foundry produced \SiN~circuits with optical loss as low as 4~dB/m. Symmetric double stripe (SDS) \SiN~supports enhanced SBS through weakly-guided pressure waves, with low optical loss and excellent power handling enabling fibre-to-fibre net gain up to 3.4~dB. We demonstrate devices and systems with high performance, including a compact Brillouin laser with up to 158~mW on-chip output power in a packaged chip, versatile Brillouin-based microwave signal processing and Brillouin dynamic gratings. This work bridges high-performance integrated Brillouin photonics with a scalable foundry process and packaging of electrical and optical interfaces. 

\end{abstract}

\maketitle

\section{Introduction}

Stimulated Brillouin scattering (SBS) is a nonlinear optical interaction between photons and GHz phonons, providing narrowband gain set by the phonon lifetime (typ. 10-100 MHz) \cite{BoydNonlinearOptics}, across orders of magnitude in optical wavelength \cite{Chauhan2021, Ko2025}. By coupling optical and acoustic fields with vastly different dissipation rates, stimulated Brillouin lasers (SBLs) exhibit phase noise filtering \cite{Debut2000, Yuan2020}, enabling the integration of narrow linewidth lasers \cite{Kabakova2013, Otterstrom2018, Gundavarapu2019, Liu2025} and RF sources \cite{Merklein2016, Ye2025, Egbert2025}. Brillouin scattering is also used as a key component in microwave photonics \cite{Eggleton2019, Liu2019, Gertler2022, Zhou2024, Ye2025}, gyroscopes \cite{Lai2020, Ge2026} and optical clocks \cite{Loh2020, Martin2026}, and shows promise for integrated optical spectrum analysis \cite{Wang2026} and photonic neural networks \cite{Slinkov2025}.

Silicon nitride (Si\textsubscript{3}N\textsubscript{4}) is a prominent platform for photonic integration, enabling ultra-low optical loss \cite{Bauters2011}, thermal tuning and III-V heterogeneous integration \cite{Xiang2022} but features negligible intrinsic SBS gain \cite{Gyger2020}. Fibre-like gain generated in the silica cladding is typically used for Brillouin lasers \cite{Liu2025}, relying on sub-dB/m losses. While successful, low confinement is needed to reach exceptionally high Q-factors. This increases bend radii to prevent radiation loss, requires specialised production lines and limits integration density.

\begin{figure*}[t!]
\centering
\includegraphics[width=\linewidth]{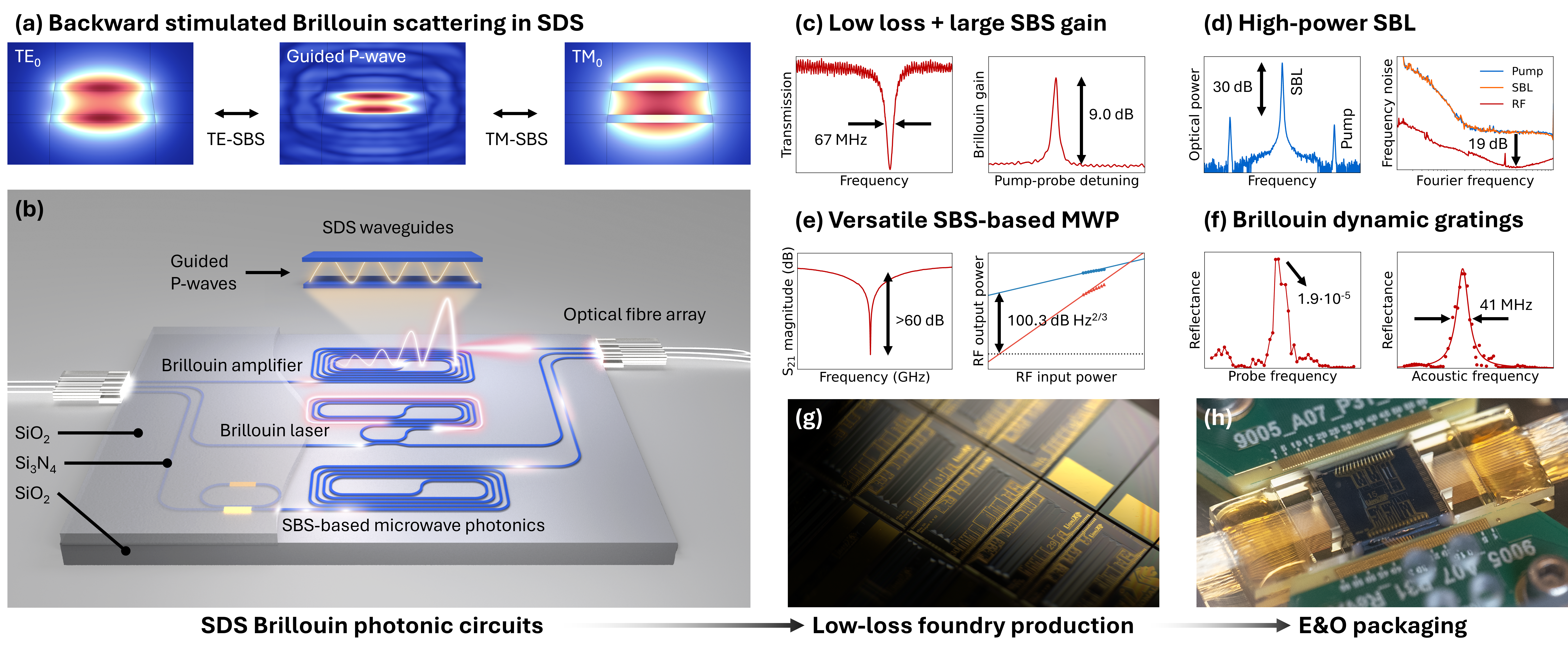}
\caption{\textbf{Scalable Brillouin photonics in symmetric double-stripe (SDS) \SiN.} \textbf{(a)} Optical and acoustic fields involved in TE- and TM-polarised, backward Brillouin scattering in SDS waveguides, mediated by guided pressure waves. \textbf{(b)} Concept illustration of Brillouin photonics in SDS, with the inset showing the SBS-active waveguide element. Standard fibre arrays interface with on-chip zero-change SBS devices, including amplifier waveguides, Brillouin lasers and Brillouin-based reconfigurable microwave processor. Top cladding removed only for visual clarity. \textbf{(c)} Low loss fabrication enabling 67~MHz TE resonator linewidth (6.7~dB/m loss) and large SBS gain in TM. \textbf{(d)} Compact, high-power (158~mW) Brillouin lasers with high side-mode suppression and RF linewidth reduction. \textbf{(e)} SBS-enhanced microwave notch filtering with high link performance. \textbf{(f)} TM-pumped Brillouin dynamic gratings with narrowband TE reflectance. \textbf{(g)} Partial view of a wafer comprising SDS \SiN~Brillouin photonic chips fabricated with a foundry process. \textbf{(h)} Electrical and optical packaging of a Brillouin laser chip.}
\label{fig1}
\end{figure*}

In this work we combine scalability with performance through zero-change SBS integration in a foundry \SiN~platform. Symmetric double-stripe (SDS) \SiN~has recently been shown to feature enhanced gain coefficients up to 4x compared to dilute \SiN~\cite{Botter2022} while maintaining low propagation loss and compactness \cite{Roeloffzen2018}. In SDS waveguides, the backwards Brillouin interaction is mediated by weakly-guided pressure waves (P-waves) between two \SiN~strips of equal thickness (Fig.~\ref{fig1}~(a)). Acoustic confinement can be controlled by waveguide width \cite{Botter2022}. Figure~\ref{fig1}~(b) shows an overview of Brillouin photonics integration enabled by SDS \SiN.

Here, we report on a range of Brillouin devices created in SDS \SiN~made with a foundry process (LioniX Int.), designed to operate at 1550 nm. We investigate the Brillouin gain spectrum of the SDS waveguides in both polarisations, validating theoretical modelling with experiments. Exploiting the excellent power handling (up to 6.3~W continuous wave (cw) on-chip power) and low loss ($4-7$~dB/m) of SDS waveguides, we demonstrate an SBS amplifier with fibre-to-fibre net gain up to 3.4~dB in the TM configuration (Fig.~\ref{fig1}~(c)). By embedding the SBS gain in a low-loss resonator, we show a compact ($0.7~\mathrm{mm}^2$) stimulated Brillouin laser (SBL) with record on-chip Stokes power of 158~mW in an electrical and optical (E\&O) packaged device (Fig.~\ref{fig1}~(d, h)). We employ the SBS response in a microwave photonics (MWP) system capable of time delay up to 5~ns, wideband phase shifting by $110\degree$ and notch filtering with 3.7~dB link gain and 17.8~dB noise figure (Fig.~\ref{fig1}~(e)). Finally we generate Brillouin dynamic gratings with up to $1.9\cdot10^{-5}$ reflectance by TM pumping (Fig.~\ref{fig1}~(f)). These results achieved using foundry \SiN~(Fig.~\ref{fig1}~(g)) underscore the readiness of system-level SBS integration, maintaining performance while benefitting stability, ease of operation and readiness for co-packaged optics (CPO) and other photonic-electronic integration strategies (Fig.~\ref{fig1}~(h)).

\section{Results}

\subsection{Polarisation-dependent Brillouin gain}

\begin{figure*}[t!]
\centering
\includegraphics[width=\linewidth]{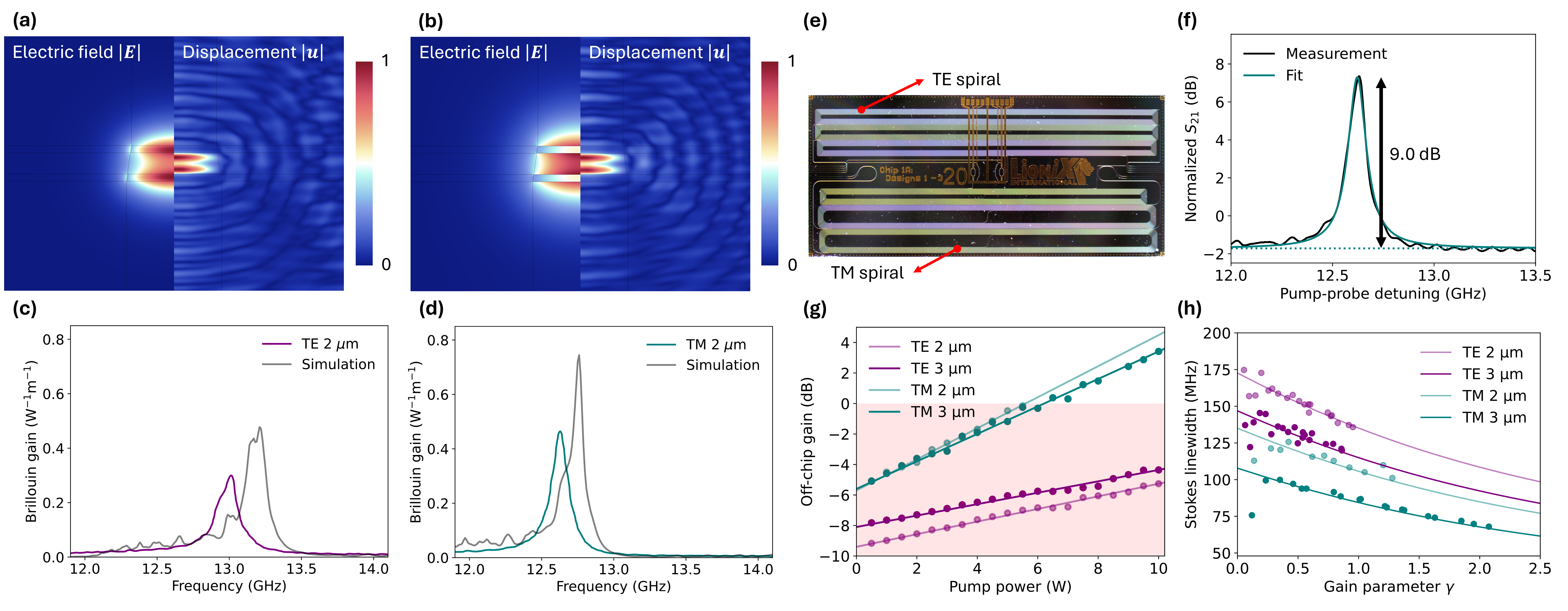}
\caption{\textbf{Polarisation-dependent Brillouin gain in SDS \SiN.} \textbf{(a, b)} Simulated optical fundamental TE and TM mode (left) and corresponding displacement responses at 13.21~GHz and 12.76~GHz pump-probe detuning (right). \textbf{(c, d)} Lock-in amplifier measurement of TE- and TM-polarised Brillouin gain spectra of 2-µm-wide SDS waveguides, compared to fit-free simulations. \textbf{(e)} Brillouin amplifier chip featuring several 1-metre long spiral waveguides designed for TE or TM propagation. \textbf{(f)} Measured VNA $\mathrm{S}_{21}$, normalised by the pump-off measurement, showing the largest recorded SBS gain up to 9.0~dB using the 3-µm-wide, TM waveguide. \textbf{(g)} Amplifier performance for different widths and polarisations, showing measured fibre-to-fibre net SBS gain versus pump power, where the red shaded region indicates insertion loss. Net gain up to 3.4~dB is obtained from the data in (f). \textbf{(g)} Linewidth of the amplified Stokes emission versus gain parameter $\gamma$. The smallest $\Delta\nu_S=68$~MHz, or 63\% of the intrinsic linewidth ($\gamma = 0$) is found in the 3~µm, TM waveguide.}
\label{fig2}
\end{figure*}

We start by verifying the TE- and TM-polarised Brillouin gain spectrum of fabricated devices by comparing detailed measurements with simulations. We consider the Brillouin gain of SDS waveguides with 170~nm \SiN~layer thickness and 500~nm inter-layer spacing oxide. We simulate the gain spectrum based on the framework in \cite{Rakich2012}, with acoustic simulations accounting for geometric (radiation) and absorption loss. Several waveguide widths are simulated between 1.2~µm and 4.0~µm. We focus on 2~µm waveguide width, also used in Brillouin amplifier and MWP applications.

Figure~\ref{fig2}~(a) show the simulated fundamental TE-like optical mode and displacement response at the Brillouin gain peak, respectively. The same is shown for TM-polarised SBS in Fig.~\ref{fig2}~(c). The simulated Brillouin gain spectrum from $12-14$~GHz in the two polarisations is shown in Fig.~\ref{fig2}~(b) and (d) along with calibrated lock-in measurements. Simulations for all widths and polarisation are shown in supplement~A. The width range of 1.2~µm to 4.0~µm results in gain responses with linewidths of about 280~MHz to 90~MHz for the TE polarisation, reducing further to 40~MHz for the TM polarisation. 

We investigate two sets of 1-metre-long spiral waveguides, see the fabricated chip layout in Fig.~\ref{fig2}~(e). Waveguide widths are 2~µm and 3~µm and input tapers are designed for either TE or TM coupling to standard single-mode fibre (SMF) with expected loss around 1~dB/facet. The bending radius of TM waveguides is 300~µm to prevent radiation loss \cite{Botter2024}, TE waveguide bends have a radius of 100~µm. The measured spectra shown in Fig.~\ref{fig2}~(b) and (d) show good qualitative agreement with the simulation. For TE, we find $0.30\pm0.05~\mathrm{W}^{-1}\mathrm{m}^{-1}$ at 13.01~GHz with a line shape closely matching the simulated response. In TM this is $0.47\pm0.08~\mathrm{W}^{-1}\mathrm{m}^{-1}$ at 12.63~GHz, and the gain line is narrower, likely due to the longer acoustic wavelength being better matched to the \SiN~layer spacing. The offset in frequency shift can be attributed to uncertainties in deposited layer thicknesses, etch angle and variations in material properties like refractive index or density used in simulation. The difference between measured and simulated peak gain might indicate underestimation of acoustic loss, or spectral broadening due to waveguide inhomogeneity \cite{Wolff2016}.

\subsection{Net-gain Brillouin amplifier}

A major drawback of \SiN~is its lack of native optical gain, which is commonly mended by III-V heterogeneous integration \cite{Xiang2022}. More recently Er\textsuperscript{3+}:\SiN~waveguides emerged with $>30$~dB small-signal gain \cite{Liu2022}, but these face the same ASE limitation as fibre-based rare-earth amplifiers. SBS is a promising candidate for on-chip amplifiers, uniquely providing all-optical, reconfigurable and narrowband gain with little out-of-band ASE.

Here we use the 1-metre-long SDS waveguides to generate large Brillouin amplification in monolithic \SiN. Using a vector network analyser (VNA)-based setup, we measure SBS gain with a single-sideband modulated probe signal. Full experimental details and data processing are discussed in supplement~C. We inject a high-power pump laser counter-propagating with respect to the probe, detuned to amplify the probe sideband. The RF beat note between the probe carrier and sideband is recorded by the VNA. This setup converts SBS amplification into the RF domain by recording the $\mathrm{S}_{21}$ parameter.

Figure~\ref{fig2}~(f) shows the result with the largest amplification, in the 3~µm, TM waveguide with around 6.3~W cw power on-chip. This data is normalised by the pump-off measurement, and reveals a maximum gain of 9.0~dB from the measurement background (dashed line). The negative background level of the $\mathrm{S}_{21}$ in dB is due to drift in the experimental setup, likely increasing probe coupling loss. We determine the gain amplitude and linewidth by fitting $\mathrm{S}_{21}$ in dB with a Lorentzian profile. The linewidth extracted in this manner is the intrinsic SBS linewidth corresponding to phonon lifetime (supplement~C). 

We repeat this measurement for all waveguide samples and calculate the fibre-to-fibre gain, shown in Fig.~\ref{fig2}~(g). From the slope of these curves, we extract SBS gain coefficients of each waveguide. For the TE waveguides of 2- and 3-µm widths we obtain $0.31~\mathrm{W}^{-1}\mathrm{m}^{-1}$ and $0.25~\mathrm{W}^{-1}\mathrm{m}^{-1}$, respectively. For TM waveguides this increases to $0.55~\mathrm{W}^{-1}\mathrm{m}^{-1}$ and $0.48~\mathrm{W}^{-1}\mathrm{m}^{-1}$. These results fall within estimated error bars for all but the 3~µm, TE waveguide.

Net-gain is achieved in the 3~µm, TM waveguide for on-chip pump power above 4.1~W, overcoming the total insertion loss of 5.6~dB, see in Fig.~\ref{fig2}~(g). The largest net-gain is 3.4~dB obtained at 6.3~W pump power, the first fibre-to-fibre net SBS gain in \SiN~to the best of our knowledge. While large SBS gain typically requires specialised materials or fabrication steps, the demonstrated 9.0~dB SBS gain in zero-change SDS \SiN~presents a shift towards strong SBS realised with generic, foundry-produced photonic circuits.

\begin{figure*}[t!]
\centering
\includegraphics[width=\linewidth]{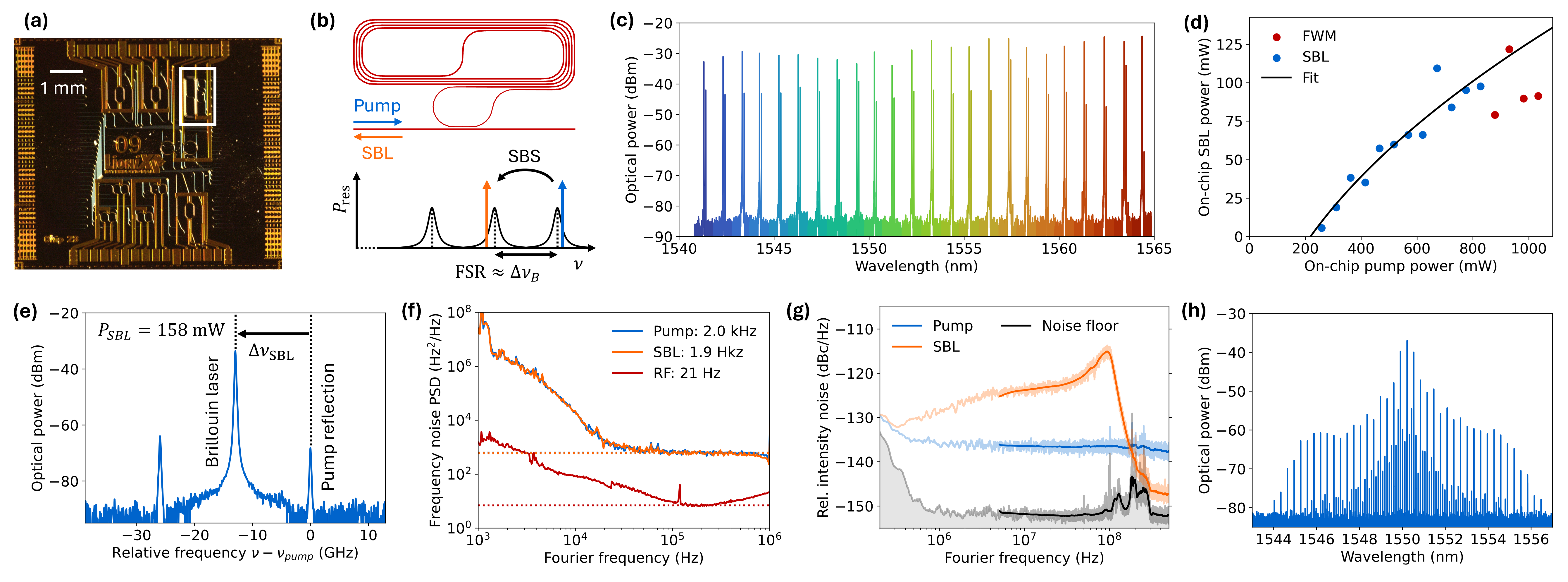}
\caption{\textbf{Compact, high-power Brillouin laser.} \textbf{(a)} Chip layout, with the Brillouin laser indicated in the white box occupying just $0.6\times1.2~\mathrm{mm}^2$. \textbf{(b)} Spiral resonator SBL schematic and operating principle. By matching the FSR to the Brillouin shift, the pump laser generates Stokes lasing in the adjacent, red-shifted resonator mode. \textbf{(c)} Tuning of the SBL from 1541~nm to 1564~nm. \textbf{(d)} Measured on-chip SBL output versus pump power near 1550~nm, revealing the threshold of $P_{\mathrm{th}}=219$~mW and slope efficiency $\mathrm{SE}=0.26$ from fitting. The on-chip Stokes power is limited by frequency comb generation, and the highest conversion efficiency is 16.3\% pumped at 672~mW. \textbf{(e)} High-power Brillouin laser at $\nu_{\mathrm{pump}}=193.419$~THz generated in the packaged laser chip (Fig.~\ref{fig1}~(h)), with 158~mW on-chip Stokes power and $30$~dB side-mode suppression. \textbf{(f)} Measured frequency noise spectra of the pump and SBL lines and their RF beat note, revealing over $90\times$ linewidth narrowing. The 21~Hz intrinsic linewidth of the RF beat note is an upper bound, limited by the measurement noise floor. \textbf{(g)} Measured relative intensity noise of the pump (before and after amplification) and Brillouin laser. At high Fourier frequencies, the RIN of the SBL ($-147$~dBc/Hz) drops below the pump ($-138$~dBc/Hz). \textbf{(h)} Brillouin-Kerr frequency comb generated with 1.47~W pump power near 1550~nm, showing 1- and 3-FSR line spacings centred on the SBL.}
\label{fig3}
\end{figure*}

Another benefit of the SBS process comes from its exponential nature, which results in a linewidth narrowing effect. The linewidth of the amplified Stokes light decreases with pump power according to \cite{Marpaung2015}
\begin{equation} \label{eq:Amp:Stokes_lw}
    \Delta\nu_{S} = \Gamma_B \sqrt{ \frac{\gamma}{\ln(e^\gamma + 1) - \ln(2)} - 1 }.
\end{equation}
Here, $\Gamma_B$ is the intrinsic SBS linewidth, i.e. phonon decay rate, which decreases as a function of the gain parameter $\gamma = g_B P_{\mathrm{pump}} L_{\mathrm{eff}}$ with on-chip pump power $P_{\mathrm{pump}}$ and effective interaction length $L_{\mathrm{eff}}$  \cite{Botter2022}. 

SDS waveguides offer control over the Brillouin gain bandwidth through waveguide width, as previously shown through simulations \cite{Botter2022}. We confirm this and show further extension of the range of Stokes linewidths through the exponential narrowing effect. We calculate $\gamma$ directly from peak SBS gain (see supplement~C) and plot $\Delta\nu_{S}$ determined from the measurements, see Fig.~\ref{fig2}~(h). Note that the uncertainty increases at low $\gamma$ due to background oscillations from the modulator and waveguide FP reflections. Solid lines show the expectation from Eq.~\eqref{eq:Amp:Stokes_lw}, using the mean $\Gamma_B$ of each waveguide. The data confirm the linewidth reduction with waveguide width and show Stokes linewidth compression with $\gamma$ in line with Eq.~\eqref{eq:Amp:Stokes_lw}. SBS linewidths are lower in TM compared to TE, resulting from better acoustic confinement with lower effective index in TM. Smaller linewidths are obtained in the wider waveguides, in line with simulation and lock-in data. With the lowest Stokes linewidth of 68~MHz at a gain parameter $\gamma\approx2.1$ (corresponding to the data in Fig.~\ref{fig2}~(f)), we show SBS linewidth between $68-175$~MHz simply by tuning waveguide width and polarisation.

\subsection{High-power Brillouin laser}

Stimulated Brillouin lasers have attracted much research attention owing to their potential for ultra-narrow linewidths \cite{Debut2000, Gundavarapu2019} and microwave generation \cite{Egbert2025, Ye2025}. Here we show a Brillouin laser in SDS, generated in a compact spiral ring resonator with the FSR matched to the Brillouin shift. This device is designed for the TE polarisation and feature 200-µm bend radius, resulting in a compact device footprint of only $0.7~\mathrm{mm}^2$. The resonator is slightly under-coupled with coupling efficiency $\eta=0.41$ for the measured 6.7~dB/m loss (Fig.~\ref{fig1}~(c)). The waveguide width inside the resonator is 4~µm, only tapering to the routing waveguide width of 1.2~µm in the coupling region. The larger waveguide width is chosen for higher Brillouin gain (supplement~A), which should reduce threshold power.

Figure~\ref{fig3}~(a) shows the SBL chip with the device under test highlighted in the white box. The waveguide layout is shown in Fig.~\ref{fig3}~(b) along with the working principle. A pump laser is tuned into the cavity resonance, building up circulating power until it reaching the SBL threshold. Above threshold, a strong Stokes line is generated in the lower frequency resonance one mode number down from the pump. We start with measurements on an unpackaged chip (experimental setup in supplement~D). By tuning the pump wavelength, see Fig.~\ref{fig3}~(c), we can generate SBL across any resonance pair from 1541~nm to 1564~nm, limited by the bandwidth of the erbium-doped fibre amplifier (EDFA) used to amplify the pump laser.

To determine the performance of our SBL, we measure the output Stokes versus input pump power, shown in Fig.~\ref{fig3}~(d). Brillouin lasing appears from 258~mW on-chip pump, increasing in power roughly following the predicted square-root relationship used in the fit. From the fit we determine the threshold power $P_{\mathrm{th}} = 219$~mW and slope efficiency $\mathrm{SE} = 0.26$ near threshold. The on-chip SBL output of 110~mW is obtained with 672~mW pump power, reaching a conversion efficiency of 16.3\%, which can be improved by adapting the design to operate at high Stokes power. The output power of the Brillouin laser is further limited by the formation of frequency combs, which first appear at 879~mW pump (see supplement~D).

A key feature of Brillouin lasers is their low intrinsic noise \cite{Gundavarapu2019}, however to achieve stable operation, packaging is crucial. The zero-change platform allows for advanced packaging of the Brillouin laser including temperature stabilisation, electrical interfacing and angled-facet fibre arrays (Fig.~\ref{fig1}~(h)). This approach reduces reduces pump reflection and facilitates operation at high optical power improved long-term stability. In the packaged device, we optimise the emitted Stokes power by increasing the pump power to 1.47~W on-chip while maintaining a larger pump detuning in order to prevent comb formation. The resulting on-chip SBL output power is 158~mW, corresponding to the spectrum shown in Fig.~\ref{fig3}~(e). The pump reflection is greatly reduced compared to the unpackaged SBL, resulting in a side-mode suppression of $-30$~dB, removing the need for output filtering.

We investigate the noise properties of the packaged SBL, measuring the frequency noise of the pump and Stokes light and their RF beat note. Depending on the dissipation rates of the optical and acoustic fields, the transfer of pump phase noise to the generated Stokes or acoustic field is reduced \cite{Debut2000, Otterstrom2018}. This enables intrinsic linewidth reduction, either of the Stokes light or the RF tone generated by heterodyning the pump and Stokes waves. The latter reflects phase noise reduction of the acoustic field in the SBL. Figure~\ref{fig3}~(f) shows the measured frequency noise of the three waves in our SBL. The pump and SBL intrinsic linewidths, corresponding to the white frequency noise level, are indistinguishable at 2.0~kHz and 1.9~kHz, respectively. However, the RF beat note is significantly purified to below 21~Hz intrinsic linewidth, limited by the measurement equipment. This is consistent with the phonon lasing regime \cite{Otterstrom2018}, with at least $90\times$ acoustic linewidth reduction. This is distinct from previously demonstrated \SiN-based SBLs, which all operate in the optical lasing regime (supplement~D).

Depending on the application, we are also in the relative intensity noise (RIN), which for example forms an important contribution to the noise floor in microwave photonic links \cite{Marpaung2019}. In Fig.~\ref{fig3}~(g) we show the measured RIN of the SBL and the pump laser. The SBL shows high RIN at low offset frequencies due to its relaxation oscillation \cite{Gundavarapu2019} at around 100~MHz, but this drops off quickly at high offset frequency. Compared to the pump, the high-frequency RIN is reduced by around 9~dB.

Combining Brillouin and Kerr effects has proven useful for low-noise and turnkey soliton generation \cite{bai2021, zhang2024}. We observe Brillouin-induced Kerr frequency combs in our SBL device by pumping close to resonance with 1.47~W, see Fig.~\ref{fig3}~(h). Here, we observe the formation of a frequency comb spanning over 10~nm with a spectral envelope reminiscent of switching waves in normal dispersion resonators \cite{Lobanov2015}. The comb is centred on the Stokes emission line, which can be red-detuned with respect to its resonance depending on the mismatch between the FSR and Brillouin shift. This enables the SBL to exhibit primary comb phase-matching despite strong normal dispersion. The dominant 3-FSR line spacing suggests the comb is initiated by phase-matching of degenerate four-wave mixing of the Stokes with the pair of modes 3 resonances removed. In supplement~D we additionally show 1- and 2-FSR spaced combs in the unpackaged device and discuss potential mechanisms. SDS provides a novel platform for studying Brillouin-Kerr interactions, which could be essential for low-noise frequency combs in normal dispersion.

\subsection{Brillouin-based microwave photonics}

\begin{figure*}[t!]
\centering
\includegraphics[width=\linewidth]{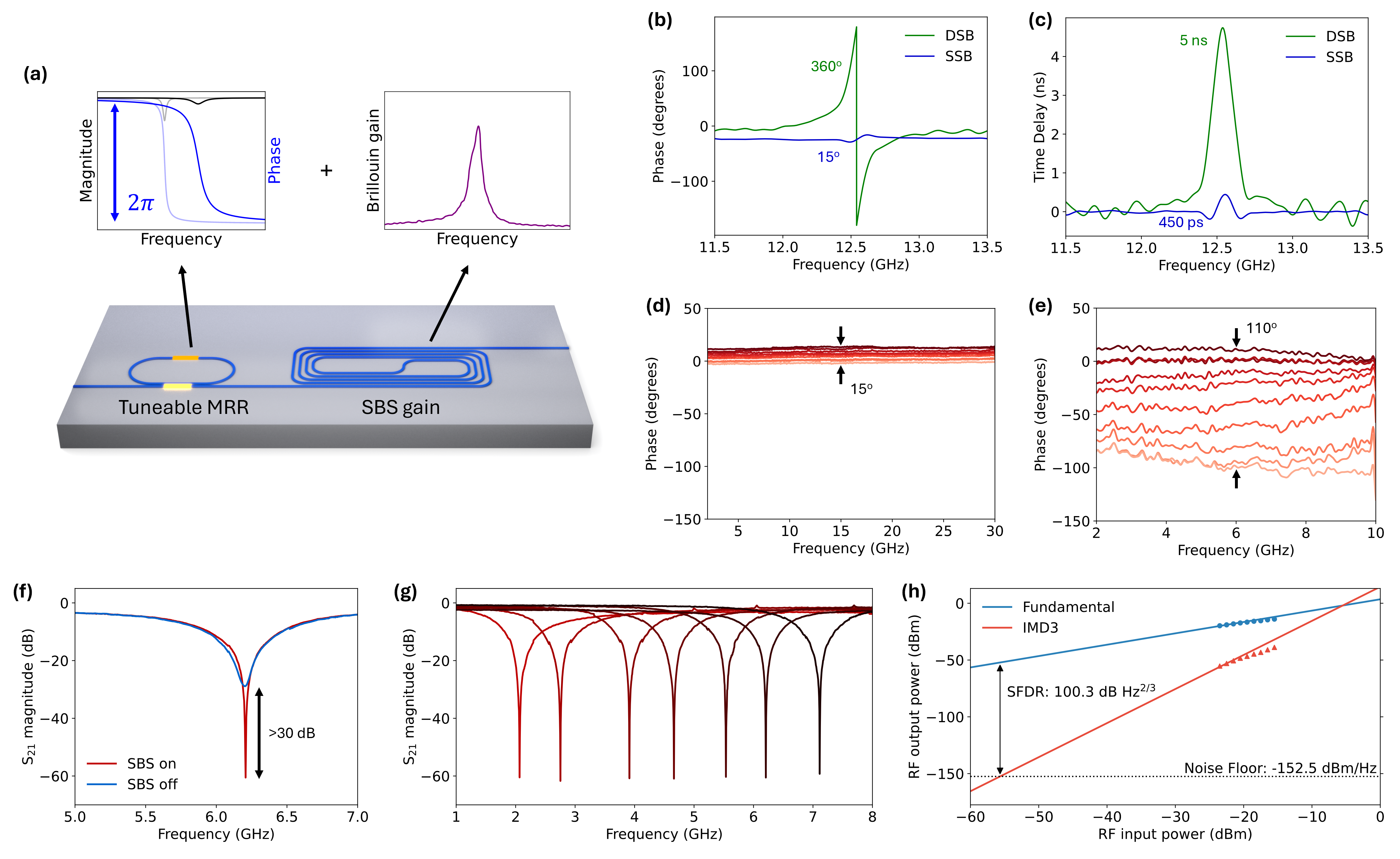}
\caption{\textbf{Versatile Brillouin microwave signal processing.} \textbf{(a)} Schematic overview of the Brillouin MWP signal processor consisting of a fully tuneable microring resonator (MRR) and SBS spiral. \textbf{(b)} Measured narrowband phase shift due to the SBS response with single sideband (SSB) and dual sideband (DSB) processing, achieving a maximum $15\degree$ and $360\degree$ phase shift, respectively. \textbf{(c)} Corresponding time delay showing enhancement from 450~ps to 5~ns using DSB processing. \textbf{(d)} Wide band phase shift by carrier processing with SSB, achieving $15\degree$ shift up to 30~GHz. \textbf{(e)} Wideband phase shifter using DSB achieving $110\degree$ with a relatively flat response up to 10~GHz. \textbf{(f)} Measured MRR notch filter without and with SBS gain, increasing the extinction to $>60$~dB. \textbf{(g)} Tuning of the notch filter central frequency over 5~GHz. \textbf{(h)} Link performance of the notch filter, revealing high link gain of 3.7~dB and a dynamic range $\mathrm{SFDR}=100.3~\mathrm{dB}~\mathrm{Hz}^{2/3}$.}
\label{fig4}
\end{figure*}

SBS has become a key component in integrated microwave photonics, providing unique benefits for reconfigurable filtering, signal generation and time delay \cite{Eggleton2019}. Integrating enhanced SBS in an SDS circuit with low-loss and mature photonic components enables an array of functionalities with high performance. We create a versatile MWP signal processor shown in Fig.~\ref{fig4}~(a), consisting of a tuneable MRR with SBS gain from a 1-metre-long, 2~µm wide TE waveguide integrated on a single chip.

We use the SBS phase response to generate both narrow and broadband phase shifting and time delay. Here, a VNA is used to drive an IQ modulator to generate the probe signal, details are found in supplement~E. This is done first using single-sideband (SSB) processing, then using dual sideband (DSB) processing to achieve RF phase and delay amplification \cite{Choudhary2017c}. In both cases, the probe upper sideband is processed with SBS and the transmitted signal is recorded through $\mathrm{S}_{21}$. The measured phase response is shown in Fig.~\ref{fig4}~(b) with single and dual sideband processing, achieving 15\degree~and a full 360\degree~phase amplitude, respectively. From the phase, we can calculate the corresponding time delay $\tau = -\frac{d\theta}{d\omega}$. Figure~\ref{fig4}~(c) shows the SBS-induced time delay of 450~ps, increasing to 5~ns.

Applications like RF beam forming require broadband phase shifting \cite{Liu2018}, which can be achieved in our system by switching to carrier processing using SBS. The pump laser is detuned by a frequency $\Omega$ roughly equal to the SBS shift up in frequency from the signal carrier, so slight adjustment of $\Omega$ imparts a linear change in phase. Figure~\ref{fig4}~(d) shows the wideband phase shifter with SSB processing, achieving 15\degree~phase shift up to 30~GHz limited by the RF equipment. The SSB scheme uses the lower sideband, since it does not overlap with the pump line. Figure~\ref{fig4}~(e) shows the result in the DSB configuration, where the interference between lower and upper sidebands amplifies the phase shift to around 110\degree up to 10~GHz. This comes at the cost of amplitude variations, increasing from 1~dB to around 14~dB (supplement~F), as well as continuous bandwidth, as the pump overlaps with the upper sideband at the SBS shift.

For many MWP systems a key challenge is the filtering of unwanted signals without degrading link performance \cite{Marpaung2013}. We combine the SBS gain with a tuneable ring resonator on the same chip, to generate a high-performance notch filter based on local destructive interference. In our system, two sideband with equal phase are generated using an intensity modulator. The ring resonator is programmed to the over-coupled regime, generating a $\pi$ phase shift on resonance, used to process the upper sideband. A pump laser generating Brillouin gain is now detuned such that the gain peak overlaps precisely with the ring response, compensating power loss from the resonator response while adding zero phase at the gain centre. The lower and upper sidebands now destructively interfere only at the $\pi$ phase shift, resulting in narrowband notch filtering with high extinction.

\begin{figure*}[t!]
\centering
\includegraphics[width=\linewidth]{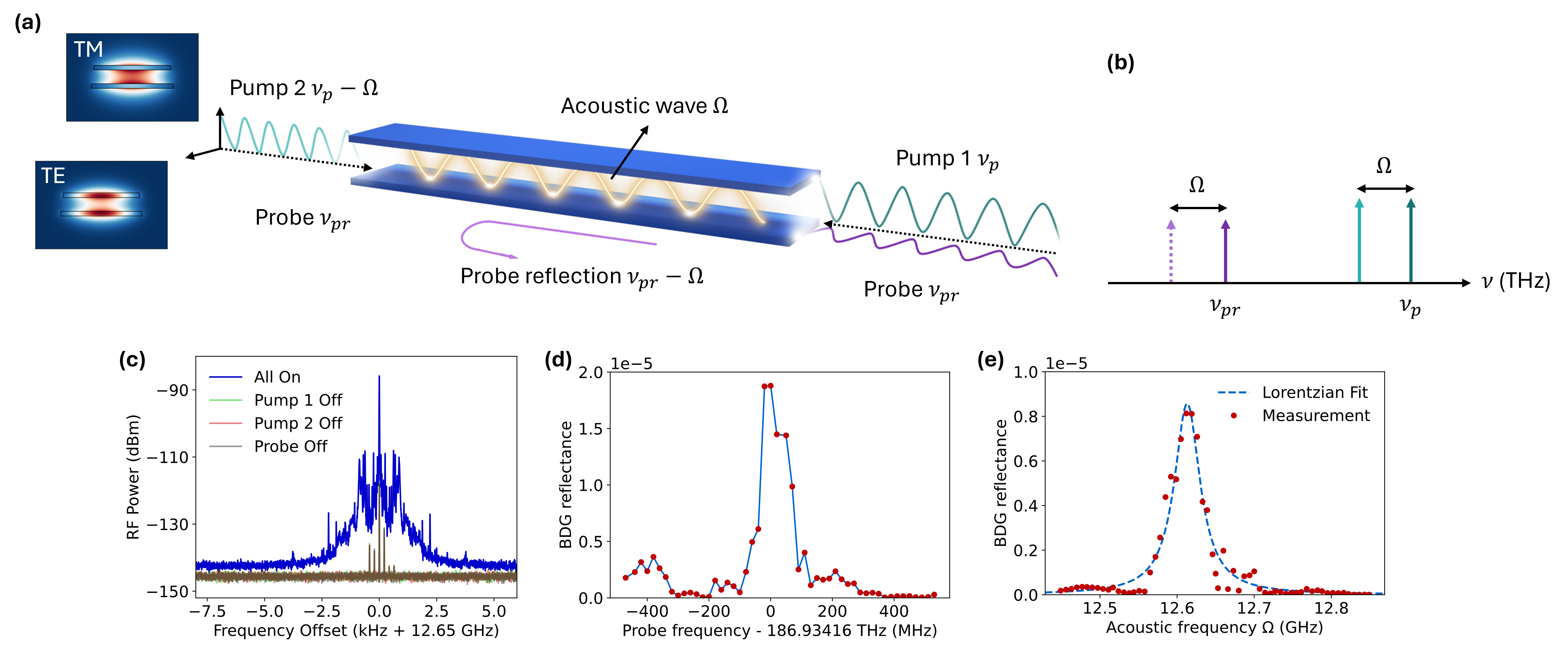}
\caption{\textbf{Brillouin dynamic gratings.} \textbf{(a)} Principle of BDG in SDS waveguides. Two pump beams around $\nu_p$ are injected in counter-propagation directions in the TM mode (intensity profile in inset), with pump 2 red-shifted by the acoustic frequency $\Omega$. This sets up a travelling acoustic wave, creating a moving index grating which is probed in the TE-polarisation at the phase-matched frequency $\nu_{pr}$. \textbf{(b)} Spectral illustration of the pump and probe optical signals. \textbf{(c)} Measured beat note between the probe reference and BDG-induced reflection signal, with a central frequency $\Omega=12.65$~GHz. The reflectance is only present when all three lasers are turned on, confirming that the signal originates from BDG. Symmetric broadening of the central reflectance signal is attributed to setup-induced phase modulation of the reference and signal waves. \textbf{(d)} Probe frequency tuning reveals a 100~MHz grating linewidth and maximum BDG reflectance of $1.9\cdot10^{-5}$. \textbf{(e)} Electronic tuning of the acoustic frequency of the BDG results in a narrow, 41~MHz linewidth reflectance for a fixed probe frequency.}
\label{fig5}
\end{figure*}

Figure~\ref{fig4}~(f) shows the RF notch filter generated by the MRR only, and the MRR + SBS combination. The filter without SBS is limited in extinction to less than 30~dB, due to sideband imbalance caused by the MRR absorption. The $-3$~dB bandwidth of the notch is 1.1~GHz, limited by to the large coupling linewidth needed for low extinction in the optical domain. The MRR + SBS filter achieves over 60~dB extinction, with minimal reduction in 3-dB linewidth. In Fig.~\ref{fig4}~(g), the central frequency of the notch filter is tuned over a 5~GHz range by shifting the pump laser and MRR resonance frequency. High performance is maintained across this bandwidth, which can be extended to a large fraction of the resonator FSR. 

We analyse the link performance of the SBS notch filter by measuring link gain (G), in-band spurious-free dynamic range (SFDR) due to third-order intermodulation distortion (IMD3) and the noise figure (NF). Low-biasing is used to increase the NF and SFDR \cite{Liu2018b} (supplement~F). The link performance is measured using a dual-tone test signal in the filter passband, where we record the fundamental (signal) power and IMD3 for a range of RF input powers, shown in Fig.~\ref{fig4}~(h). The offset between the expected slopes of 1 and 3 for the fundamental and IMD3 are due to the modulator and EDFA, reducing link gain at high RF power. To circumvent this, we only fit the three data points with lowest RF input power. With the measured noise floor of $-152.5$~dBm/Hz, we calculate the $\mathrm{SFDR} = 100.3~\mathrm{dB}~\mathrm{Hz}^{2/3}$. With the link gain $\mathrm{G}=3.7$~dB at $-23.5$~dBm RF input power, we obtain a low noise figure $\mathrm{NF} = 17.8$~dB. Low NF is achieved through low optical loss, circumventing the need for signal amplification after the chip. This underscores the unique benefit of SDS circuits integrating strong SBS in a low-loss, foundry \SiN~platform capable of performance and signal-to-noise required for real-world deployment \cite{Marpaung2013}.

\subsection{Brillouin dynamic grating}

Brillouin dynamic grating (BDGs) provide another unique route for reconfigurable, all-optical filtering. BDGs are a type of Brillouin-mediated four wave mixing where two pump lasers (frequencies $\nu_\mathrm{p}$ and $\nu_\mathrm{p}-\Omega$) generate an acoustic wave ($\Omega$), see Fig.~\ref{fig5}~(a). The acoustic wave acts as a moving index grating, which scatters the light of a probe laser in a different optical mode \cite{Song2008}. For high reflectance, the SBS process of the probe should be phase-matched to the pump-induced acoustic grating. 

While BDG has been observed in SDS in our previous work \cite{Botter2024}, this demonstration was severely limited by TM bend loss and higher propagation loss. Here, we demonstrate orders of magnitude improvements in BDG performance, enabled by increased gain in TM-SBS and simultaneous low-loss propagation of TE and TM modes. We investigate BDG in the 1-metre long, 3-µm-wide TM waveguide. Using the TM mode to set up the grating, and probing in the TE mode, results in the phase-matching condition
\begin{equation} \label{eq:bdg_phasematching}
    \lambda_{\mathrm{pr}} \approx \frac{n_{\mathrm{eff,TE}}}{n_{\mathrm{eff,TM}}} \lambda_{\mathrm{p}}.
\end{equation}
By simulating the effective index of the TE mode (1.565 at 1542~nm pump wavelength), BDG reflectance in TM is expected at $\lambda_{\mathrm{pr}}\approx1603.7$~nm or 187.0~THz from the TM effective index of 1.503. 

Figure~\ref{fig5}~(a) shows the schematic setup for BDG in the SDS waveguides, with the spectrum of the four fields in Fig.~\ref{fig5}~(b). To measure BDG, a laser at 1542~nm is used to generate the pump light (pump 1 and 2). Pump 2 is created from the carrier using a modulator and optical filter, so that the acoustic frequency is electronically controlled. Pump 1 and 2 are amplified to 2~W and 0.5~W, respectively, and injected into the SDS waveguide in the TM polarisation in counter-propagating directions. The probe laser is injected in the TE-mode, co-propagating with pump 1. The BDG reflection is extracted from the chip using a circulator and pump filter, and recorded using an electrical spectrum analyser by heterodyning with a reference probe signal. Full experimental details are found in supplement~G. Figure~\ref{fig5}~(c) shows an example of the electrical spectrum, with all lasers turned on, and pump/probe-off measurements. We see a clear signal only with all lasers turned, at the frequency $\Omega$ set by the signal generator. We attribute the symmetric sidebands in this spectrum to phase modulations of the reflected probe signal.

A unique benefit of BDG is its versatility in reflection bandwidth, behaving like a distributed index grating \cite{Dong2010} rather than being fixed by the linewidth of the SBS process used to generate it. We can estimate the BDG length using the weak, uniform grating approximation \cite{Dong2010}:
\begin{equation} \label{eq:bdg_bandwidth}
    \Delta\nu_{\mathrm{BDG}} = \frac{c}{2 n_{\mathrm{eff,pr}} L_{\mathrm{BDG}}}.
\end{equation}
We investigate the spectral resolution of the BDG by recording its reflectance while tuning the probe frequency. The extracted BDG bandwidth is $\Delta\nu_{\mathrm{BDG}}=100$~MHz, see Fig.~\ref{fig5}~(d). This corresponds to a grating length $ L_{\mathrm{BDG}}\approx0.95$~m, much longer than the cm-scale grating previously observed \cite{Botter2024}. The spectral shape of the BDG reflectance depends on the local birefringence \cite{Dong2014}, which could vary in our waveguide sample explaining the side lobes in Fig.~\ref{fig5}~(d). The maximum BDG reflectance of $1.9\cdot10^{-5}$ is an improvement by over 6 orders of magnitude compared to our previous work in SDS waveguides \cite{Botter2024}.

Another tuning mechanism for the BDG reflectance is electronically tuning the acoustic wave frequency $\Omega$. This has a two-fold effect, changing both the detuning in the pump SBS process and the probe phase-matching to the dynamic grating. In Fig.~\ref{fig5}~(e) we plot BDG reflectance for fixed pump and probe frequencies, as a function of $\Omega$, revealing a roughly Lorentzian profile with 41~MHz FWHM.

The demonstrated BDG shows that metre-scale, all-optical gratings can be achieved in standard SDS \SiN~waveguides, which could be applied to reconfigurable single- or multi-tap filtering \cite{Chin2012} and variable delay \cite{Sancho2002}. Alternatively, BDG might be used for distributed sensing \cite{Song2010} or high-resolution optical spectrum analysis \cite{Dong2014}. In our system, the large birefringence means the probe reflectance bandwidth is very different from the pump laser bandwidth. With around 4.1\% simulated birefringence, a TM-polarised pump in the C-band ($1535-1565$~nm) results in the phase-matched TE probe wavelength varying from $1598-1630$~nm according to Eq.~\eqref{eq:bdg_phasematching}. The non-overlapping pump and signal bands facilitate spectral pump filtering, removing the need for polarisation-based filtering used in fibre BDG.

\section{Discussion}

In conclusion, we use SDS \SiN~waveguides to achieve an array of Brillouin-based applications with high performance. Using the TM mode enables Brillouin amplification comparable to or higher than other scalable platforms \cite{Lei2024, Ye2025, Klaver2026} and fibre-to-fibre net gain for the first time in a \SiN~waveguide. Leveraging mature SDS integration technology, we demonstrate a compact Brillouin laser with record on-chip power. The zero-change SBS platform facilitates integration with standardised components, which we use to realise a versatile SBS-based microwave signal processor with high performance. Finally, we generate metre-scale Brillouin dynamic gratings with over a factor $10^6$ improvement in reflectance compared to our previous work \cite{Botter2024}. 

With rapid developments in photonic integration and growth in number of Brillouin-active platforms, \SiN~remains the platform of choice for advanced programmable circuits with high optical power, linearity and low noise figure. Compared to other scalable Brillouin photonics platforms such as TFLN \cite{Ye2025}, SDS benefits from a mature technology platform enabling efficient fibre-chip coupling and stability afforded by advanced packaging. We envision Brillouin-based microwave processing in SDS combined with techniques such as modulation transformation \cite{Daulay2022} and linearisation \cite{Liu2020} to reach performance and integration level required for next-generation communications systems \cite{Marpaung2019}. SDS also readily enables complex, tuneable resonator design, which can lower the SBL threshold through engineered mode splitting \cite{Ji2024} or using the TM mode, with potential for diode laser pumping. We also expect future work to continue exploring Brillouin-Kerr combs in SDS with the potential for low-noise switching waves \cite{Bunel2025}.

The results encompassed in this work are at a turning point in on-chip SBS research, moving from specialised materials and processes to generic integration platforms. The current embodiment of SDS Brillouin photonics shows significant progress in key metrics like SBS gain, lasing power and MWP link performance. Nevertheless, further improvements are required for truly scalable integration with III-V hybrid integration and CMOS-compatibility. We also expect further improvement by cross-section optimisation \cite{Botter2022} and lower optical loss to reach Brillouin amplification beyond the 10-dB level. This could unlock demanding applications like frequency-selective amplification for small signal recovery, like individual comb lines used in coherent communication \cite{Choudhary2017b, Corcoran2025}. The addition of packaging with stable E\&O interfaces will accelerate SBS integration with mature integrated photonic systems, shrinking the gap between research and real-world deployment.

\section{Methods}

\noindent
\textbf{Fabrication and packaging}

Details of the waveguide and cladding depositions are found in our previous work \cite{Botter2022}. In this work we perform waveguide patterning using deep-ultraviolet stepper lithography followed by reactive ion etching. Input tapers are designed separately for TE and TM waveguides for fibre-chip coupling loss below 1~dB.

We demonstrate advanced packaging of the SBL chip to address thermal and optical coupling stability, minimising long-term drift and improving ease of operation. Packaging of the Brillouin laser chip was carried out by Phix Bv, using angled-facet fibre arrays polished at $8\degree$ in the vertical direction to suppress pump reflections. Fibre arrays are mounted without glue in the optical path to allow high power operation. \\

\noindent
\textbf{Brillouin laser noise characterisation}

Setups used for optical and RF linewidth measurements are discussed in detail in supplement~D. The linewidth of the acoustic field in the SBL is reflected in the RF beat note generated by heterodyning the pump and Stokes waves \cite{Otterstrom2018}. We use a correlated self-heterodyne method \cite{Yuan2022} to measure the optical phase noise, where no pump filter is required the packaged laser. For the frequency noise of the RF beat note we use a commercial phase noise analyser (Keysight CXA SignalAnalyzer N9000B).

For RIN measurements, the setup is adjusted by adding a bias tee and RF amplifier between the photodiode and ESA, removing dc photocurrent to record intensity fluctuations with high signal-to-noise. The voltage on the bias tee needed to cancel dc photocurrent is used to calculate the optical power on the photodiode using its responsivity. The optical power of both the pump and SBL is set to 6~dBm to ensure direct comparison.

\bigskip
\section*{Author Contribution}
\label{sec:four}

Conceptualisation: D.M., R.B. and Y.K.
Methodology: R.M., Y.K., A.T.I., A.K., J.N., J.B., A.Mi., O.D., P.v.d.S.
Chip design: R.B., O.D., Y.K., A.Me., E.K.
Software: R.M., Y.K., A.T.I., J.N., J.B.
Resources: A.Me. and E.K.
Writing (original draft): R.M., A.T.I., J.N., D.M.
Writing (review and editing): R.M., A.T.I., J.N., J.B., A.K., R.B., O.D., P.v.d.S., A.Me., E.K., Y.K., D.M.
Funding acquisition and supervision: D.M.

\begin{acknowledgments}
The authors acknowledge funding from the European Research Council Consolidator Grant (101043229 TRIFFIC), Nederlandse Organisatie voor Wetenschappelijk Onderzoek (NWO) Start Up (740.018.021).

\end{acknowledgments}

The authors thank Redlef Braamhaar for preparing photographs and thank Frauwelt Photonics Bv for artist illustrations.

The authors acknowledge chip packaging by PhiX Bv.

\section*{Disclosures}

The authors declare no conflict of interest.

\section*{Data Availability}
The data of this study are available from the corresponding authors upon reasonable request.

\bibliographystyle{IEEEtran}
\bibliography{library}

\end{document}